# Phonon anharmonicity and thermal conductivity of two-dimensional van der Waals materials: A review

Xuefei Yan[1], Bowen Wang[1], Yulong Hai[1], Devesh R. Kripalani[2], Qingqing Ke[3,4,*] and Yongqing Cai[1,*]

[1]*Joint Key Laboratory of the Ministry of Education, Institute of Applied Physics and Materials Engineering, University of Macau, China*

[2]*School of Mechanical and Aerospace Engineering, Nanyang Technological University, 50 Nanyang Avenue, Singapore 639798, Singapore*

[3]*School of Microelectronics Science and Technology, Sun Yat-sen University, Zhuhai 519082, China*

[4]*Guangdong Provincial Key Laboratory of Optoelectronic Information Processing Chips and Systems, Sun Yat-Sen University, Zhuhai 519082, China*

E-mail: keqingq@mail.sysu.edu.cn; yongqingcai@um.edu.mo

**Abstract:** Two-dimensional (2D) van der Waals (vdW) materials have extraordinary thermal properties due to the effect of quantum confinement, making them promising for thermoelectric energy conversion and thermal management in microelectronic devices. In this review, the mechanism of phonon anharmonicity originating from three- and four-phonon interactions is derived. The phonon anharmonicity of 2D vdW materials, involving the Grüneisen parameter, phonon lifetime, and thermal conductivity, is summarized and derived in detail. The size-dependent thermal conductivity of representative 2D vdW materials is discussed experimentally and theoretically. This review will present fundamental and advanced knowledge on how to evaluate the phonon anharmonicity in 2D vdW materials, which will aid the design of new structures and materials for applications related to energy transfer and conversion.

## I. Introduction

Van der Waals (vdW) materials, held together by weak electrostatic forces, are named after Dutch scientist Johannes Diderik Van der Waals who aimed to gain insight into the molecular interactions that lead to gas nonideality. Great efforts have been made to investigate the mechanical, chemical, and thermal properties of various bulk vdW materials. Incidentally, as early as 1947, Wallace had predicted that in the extreme-thickness limit of one or a few molecules, vdW materials will exhibit very different properties from their bulk[1]. However, this speculation remained unnoticed until graphene was successfully isolated in 2004[2]. Graphene, as a typical two-dimensional (2D) vdW material, is an atomically thin sheet of atoms arranged in a honeycomb structure. Monolayer graphene behaves very differently from its bulk progenitors; its exotic properties include the quantum Hall effect[3], ultra-high specific surface area (around 2630 $m^2$/g)[4], and high transmittance (up to 97.7%)[5], among others. Following the emergence of monolayer graphene, other similar layered materials were discovered and extensively investigated, spanning the complete spectrum of electronic properties, as shown in Fig. 1.

According to the thousands of articles published on 2D vdW materials over the past 20 years, thickness engineering is an effective way to modulate the performance and behavior of vdW materials due to quantum confinement and boundary effects. For instance, the high anisotropy of monolayer phosphorene[6], the large (1.9 eV) direct band gap in monolayer $MoS_2$[7-9], and the excellent ductility of atomically thin vdW materials[10] are just typical examples of abnormal behaviors in the atomic limit. Many review papers have been published on the new physical properties associated with thickness reduction, particularly for graphene[11-13], transition metal dichalcogenides (TMDs)[14-17], and black phosphorus[14, 18-20]. Therefore, those aspects are not discussed in detail here.

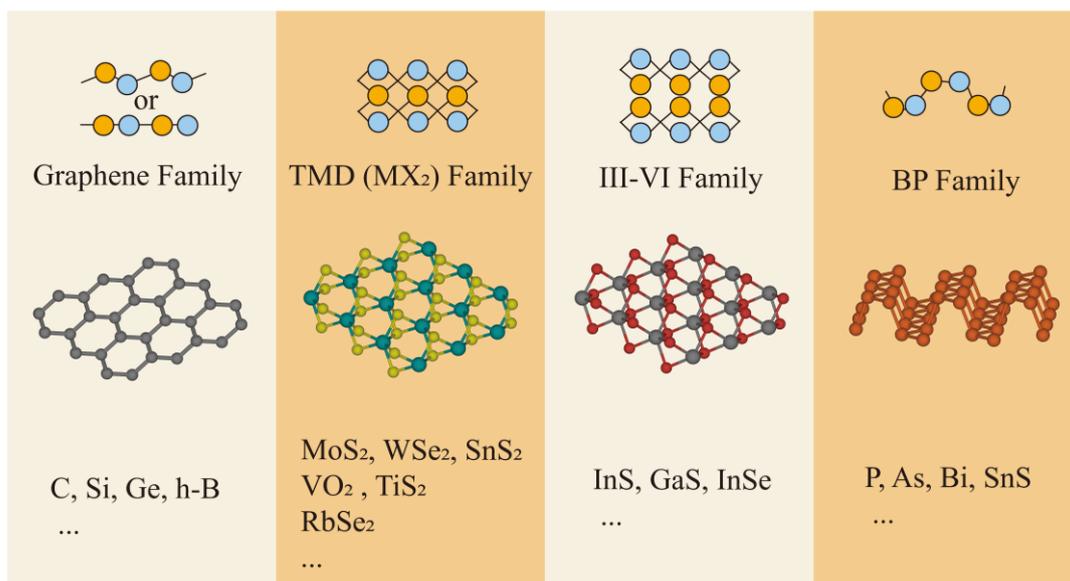

**Fig. 1** The realm of 2D vdW materials including graphene and its analogues, boron nitride (BN), black phosphorus (BP), the III-VI family, and TMDs which span the full range of electronic properties. The atomic models given at the top show the materials' cross-sectional structures—most of them are not intrinsically planar.

Besides the electronic analysis, the thermal properties of 2D vdW materials have also been of ever-increasing interest, as they not only determine the performance and reliability of nanodevices, but also govern the efficiency of thermoelectric energy conversion and thermal dissipation in microelectronic devices[21-23]. To date, the thermal properties of various 2D vdW materials, such as graphene[24-26], stanene[27, 28], and monolayer TMDs[29-34], have been widely investigated. Moreover, numerous theoretical calculations and experimental work have been conducted to understand phonon scattering processes, since the related mechanisms are critical to tailoring the thermal properties of materials. Lattice thermal resistance is mainly dictated by phonon scattering, which can be rooted in extrinsic factors (such as isotopic disorder, atomic defects, and the finite-size of crystals, *etc*.) or related to intrinsic ones. Information on the degree of phonon scattering can be obtained from infrared (IR) or Raman measurements of linewidths and line shifts of phonon modes[35]. For instance, the shift of the harmonic phonon frequency with temperature, directly accessible from experimental measurement, is rooted in the phonon-phonon interaction[35-37]. The intrinsic phonon linewidth is governed by electron-phonon[38] and anharmonic phonon-phonon interactions[39], and the latter usually makes a dominant contribution. Such anharmonic phenomena of vdW 2D materials and heterostructures are therefore under intense scrutiny. Inspired by the above, this paper reviews the mechanisms of phonon scattering including three- and four-phonon interactions and lattice thermal conductivity of 2D vdW materials.

**II. Phonon Anharmonicity**

The phonon scattering process, an intrinsic scattering mechanism responsible for the lattice thermal conductivity, plays a significant role in affecting the thermal conduction in insulators and semiconductors around and above room temperature. Particularly, by using the selection rules that allow and prohibit scattering processes of phonons in a given material, it is reasonable to establish the thermal transport properties of the material by examining the coupling of phonons. The intrinsic phonon scattering is mainly induced by intrinsic anharmonic phonon-phonon scattering which can be decomposed as three-phonon, four-phonon, and even higher-order phonon scattering processes. Various phonon scattering processes involving three-phonon and four-phonon coupling are displayed in Fig. 2.

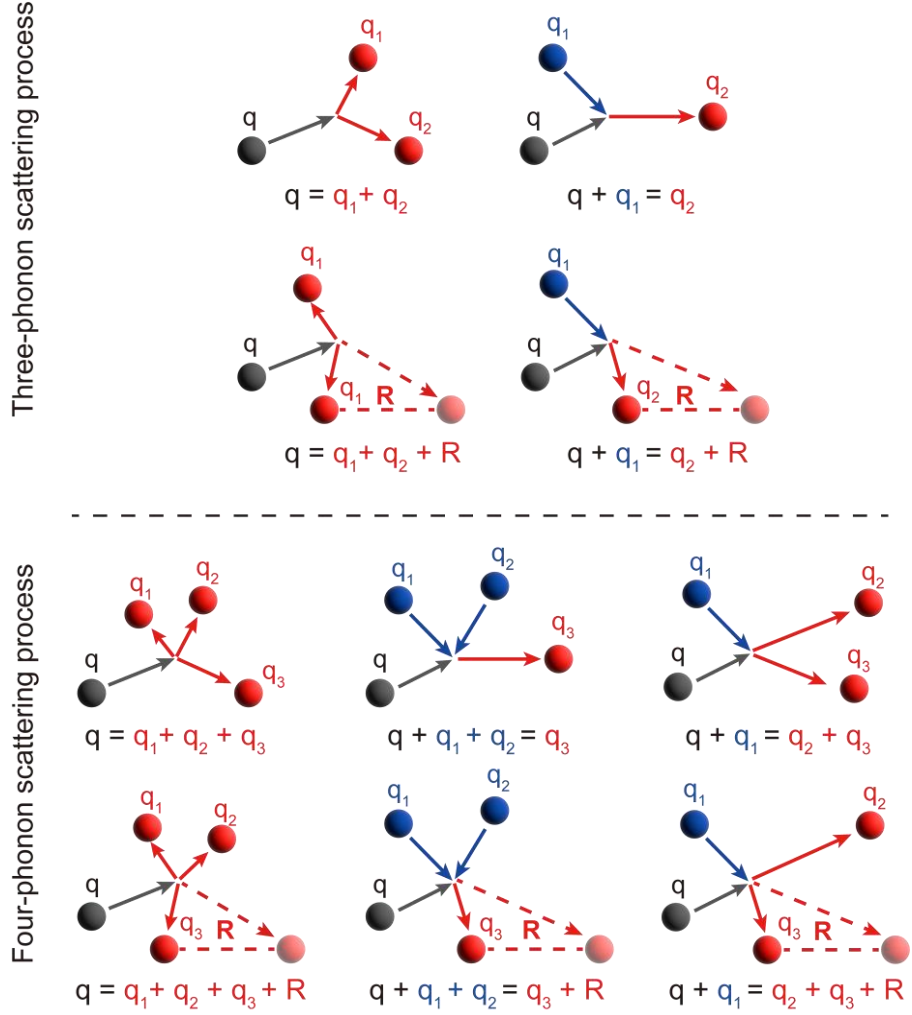

**Fig. 2** Schematic diagram of three- and four-phonon scattering processes. The phonon momentum is given by $\hbar q$, where $q$ denotes the wave vector. The process of momentum conservation is a normal process. Other processes where the momentum is not conserved are known as Umklapp processes, in which the generated phonons are folded back by the reciprocal lattice vector **R**.

From perturbation theory, the lattice Hamiltonian of a crystal can be written as[40-42]

$$H = H_0 + \underbrace{H_3 + H_4 + \cdots}_{\text{Anharmonic terms}}, \tag{1}$$

where the harmonic term $H_0$ and anharmonic terms $H_i$ (i = 3, 4, …) can be given as[40]

$$H_0 = \sum_\lambda \hbar\omega_\lambda \left( c_\lambda^\dagger c_\lambda + \frac{1}{2} \right),$$

$$H_3 = \sum_{\lambda\lambda_1\lambda_2} \frac{\hbar^{\frac{3}{2}}}{2^{\frac{3}{2}} \times 3! N^{\frac{1}{2}}} \Delta_{\mathbf{q}+\mathbf{q}_1+\mathbf{q}_2,\mathbf{R}} \frac{V^{(3)}_{\lambda\lambda_1\lambda_2}}{\sqrt{\omega_\lambda \omega_{\lambda_1} \omega_{\lambda_2}}} \left( c_{-\lambda}^\dagger + c_\lambda \right)\left( c_{-\lambda_1}^\dagger + c_{\lambda_1} \right)\left( c_{-\lambda_2}^\dagger + c_{\lambda_2} \right), \quad (2)$$

$$H_4 = \sum_{\lambda\lambda_1\lambda_2\lambda_3} \frac{\hbar^2}{2^2 \times 4! N} \Delta_{\mathbf{q}+\mathbf{q}_1+\mathbf{q}_2+\mathbf{q}_3,\mathbf{R}} \frac{V^{(4)}_{\lambda\lambda_1\lambda_2\lambda_3}}{\sqrt{\omega_\lambda \omega_{\lambda_1} \omega_{\lambda_2} \omega_{\lambda_3}}} \left( c_{-\lambda}^\dagger + c_\lambda \right)\left( c_{-\lambda_1}^\dagger + c_{\lambda_1} \right)\left( c_{-\lambda_2}^\dagger + c_{\lambda_2} \right)\left( c_{-\lambda_3}^\dagger + c_{\lambda_3} \right),$$

…

Here, $H_3$ and $H_4$ represent three-phonon and four-phonon scattering processes (incidentally, around room temperature, three-phonon scattering processes are stronger than higher-order anharmonic phonon scattering processes [43]), respectively. $\hbar$ refers to the Plank constant. $\omega_\lambda$ and $\mathbf{R}$ are the angular frequency with phonon mode $\lambda$ and a reciprocal lattice vector, respectively. $N$ is the total number of $\mathbf{q}$ points. $c_\lambda (c_\lambda^\dagger)$ is an annihilation (creation) operator. The Kronecker delta $\Delta_{\mathbf{n},\mathbf{R}}$ obeys the following momentum selection rule:

$$\Delta_{\mathbf{n},\mathbf{R}} = \begin{cases} 1 & \text{if } \mathbf{n} = \mathbf{R} \\ 0 & \text{otherwise} \end{cases}, \quad \mathbf{n} = \mathbf{q}+\mathbf{q}_1+\mathbf{q}_2 / \mathbf{q}+\mathbf{q}_1+\mathbf{q}_2+\mathbf{q}_3 \quad (3)$$

Moreover, according to Ref. [40], the anharmonic phonon scattering coefficients $V^{(n)}_\lambda$ with phonon mode $\lambda$ can be obtained as

$$V^{(3)}_{\lambda\lambda_1\lambda_2} = \sum_{b,l_1b_1,l_2b_2} \sum_{\alpha\alpha_1\alpha_2} \Phi^{\alpha\alpha_1\alpha_2}_{0b,l_1b_1,l_2b_2} \frac{e^\lambda_{\alpha b} e^{\lambda_1}_{\alpha_1 b_1} e^{\lambda_2}_{\alpha_2 b_2}}{\sqrt{m_b m_{b_1} m_{b_2}}} e^{i\mathbf{q}_1 \cdot \mathbf{r}_{l_1}} e^{i\mathbf{q}_2 \cdot \mathbf{r}_{l_2}},$$

$$V^{(4)}_{\lambda\lambda_1\lambda_2\lambda_3} = \sum_{b,l_1b_1,l_2b_2,l_3b_3} \sum_{\alpha\alpha_1\alpha_2\alpha_3} \Phi^{\alpha\alpha_1\alpha_2\alpha_3}_{0b,l_1b_1,l_2b_2,l_3b_3} \frac{e^\lambda_{\alpha b} e^{\lambda_1}_{\alpha_1 b_1} e^{\lambda_2}_{\alpha_2 b_2} e^{\lambda_3}_{\alpha_3 b_3}}{\sqrt{m_b m_{b_1} m_{b_2} m_{b_3}}} e^{i\mathbf{q}_1 \cdot \mathbf{r}_{l_1}} e^{i\mathbf{q}_2 \cdot \mathbf{r}_{l_2}} e^{i\mathbf{q}_3 \cdot \mathbf{r}_{l_3}}, \quad (4)$$

…

where $e^\lambda_{ab}$ is the phonon eigenvector of phonon mode $\lambda$. $b$ and $\alpha$ denote indices of the internal atomic coordinates and Cartesian index in the unit cell ($l$) identified by the lattice vector $\mathbf{R}$, respectively. $\mathbf{r}_l$ is the position vector of the $l$-th unit cell. $m_b$ is the average atomic mass at the $b$ atom. The elements of the interatomic force constants (IFCs) matrix ($\Phi^\alpha_{lb}$) can be defined as [44, 45]

$$\Phi^{\alpha\alpha_1\alpha_2}_{0b,l_1b_1,l_2b_2} = \left.\frac{\partial^{(3)}E}{\partial r^{\alpha}_{0b}\partial r^{\alpha_1}_{l_1b}\partial r^{\alpha_2}_{l_2b}}\right|_0,$$

$$\Phi^{\alpha\alpha_1\alpha_2\alpha_3}_{0b,l_1b_1,l_2b_2,l_3b_3} = \left.\frac{\partial^{(4)}E}{\partial r^{\alpha}_{0b}\partial r^{\alpha_1}_{l_1b}\partial r^{\alpha_2}_{l_2b}\partial r^{\alpha_3}_{l_3b}}\right|_0, \quad (5)$$

...

where $E$ is the total potential energy of the crystal, and $r^{\alpha}_{lb}$ refers to a displacement along the $\alpha$ direction. To obtain the interatomic force constants, molecular dynamics (MD) simulation or first-principles calculations are usually preferred. Phonon dispersion and the anharmonic phonon effect can thus be satisfactorily described *via* first-principles calculations or MD simulation[46]. It is worth mentioning that by using the extracted force constants to solve the Boltzmann transport equation, the obtained lattice thermal conductivity generally agrees well with experimental results for various materials[47-50]. Moreover, the reader of this article will have no problem finding many reviews on the anharmonic phonon properties (including the Grüneisen parameter, phonon lifetime, and phonon frequency shift, *etc*.) and thermal conductivity of 2D vdW materials.

**A. Grüneisen parameter**

Thermal expansion is a critical factor affecting the performance of 2D devices[51, 52]. Particularly, differences in the thermal expansion coefficient between the sample and substrate could induce external strain that affects device performance and its operational lifetime. Mathematically, the effect is captured by the Grüneisen parameter[53], which is given by

$$\gamma = \left.\frac{dP}{d(E_\lambda/V)}\right|_V = \frac{\alpha v_s^2}{C_p}, \quad (6)$$

where $E_\lambda$ is the internal energy attributed to mode $\lambda$. $P$ and $V$ refer to the pressure and volume of the sample, respectively. $\alpha$ is the thermal expansion coefficient and $C_P$ denotes the heat capacity per unit mass at constant pressure. $v_s$ refers to the sound velocity. The equilibrium structure at any temperature $T$ can be obtained by minimizing the Helmholtz free energy without applying pressure [54]. In the context of physics, the Grüneisen parameter can be derived by assuming a linear dependence of the phonon frequencies on the three orthogonal cell dimensions[54, 55]. Therefore, for 2D vdW materials, the Grüneisen parameter of each phonon mode can be rewritten as[54]

$$\gamma_q = -\frac{a}{2\omega(\mathbf{q})}\frac{\partial \omega(\mathbf{q})}{\partial a}\bigg|_0. \tag{7}$$

where $\omega(\mathbf{q})$ and $a$ refer to the phonon frequency and lattice constant, respectively. By combining Eqs. (2) and (4) with Eq. (7), it is not difficult to see that the Grüneisen parameters are directly related to the Fourier change of the third-order force constant[43].

The study of Grüneisen parameters for 2D vdW materials originated in 2009[56, 57]. Kim *et al.* reported the existence of negative Grüneisen parameters in both monolayer and bilayer graphene (see Fig. 3), indicating negative thermal expansion coefficients[57]. Later, negative Grüneisen parameters were also found in black and blue phosphorene[58-61], arsenene and antimonene[61], multilayer structures stacked by vdW interactions[62], as well as 2D IV-VI compounds (GeS, GeSe, SnS, and SnSe)[63]. Physically, only phonon modes that soften with decreasing lattice constant have negative Grüneisen parameters [see Eq. (7)]. Also, the negative Grüneisen parameter corresponds to an increase in phonon frequency with increasing lattice constant. First-principles calculations have demonstrated that the negative thermal expansion in these 2D vdW materials are mostly influenced by the out-of-plane ZA/ZO mode. In other words, at low frequencies, the collective vertical motion of the atoms is strong, and the Grüneisen parameters could be negative. However, all the Grüneisen parameters are positive in the study of monolayer $MoS_2$ by Cai *et al.*[29]. Indeed, the phonon frequencies of materials generally decrease as the material thermally expands, giving rise to positive Grüneisen parameters. Although low-frequency acoustic modes produce some negative Grüneisen parameters, sometimes competing with positive ones, negative thermal expansion will primarily occur at low temperatures when only the lowest acoustic modes can be excited.

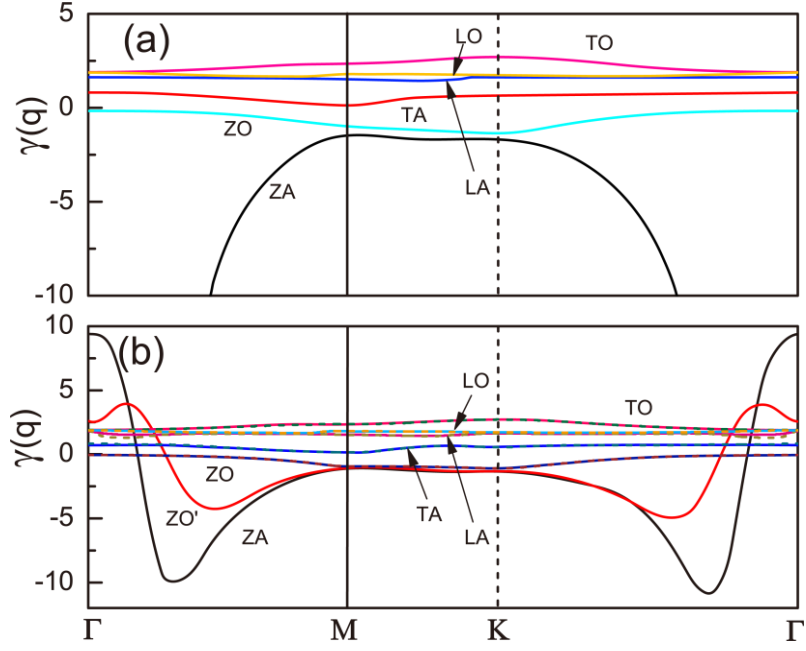

**Fig. 3** Grüneisen parameters of (a) monolayer graphene and (b) bilayer graphene (with A-B stacking) along the high-symmetry directions[57]. Reproduced with permission. Copyright (2009), American Physical Society.

**B. Phonon lifetime**

The phonon lifetime is related to the phonon energy linewidths *via* the Heisenberg uncertainty relation. Various scattering events involving phonons (such as phonon–phonon Umklapp scattering, mass-difference scattering, phonon–defect scattering, *etc*.) should be considered in the calculation of phonon lifetime, as they all serve to broaden the phonon peaks measured in constant-Q scans [40]. However, only three-phonon Umklapp scattering and mass-difference scattering are considered as intrinsic, while phonon–defect scattering can be extrinsic and eliminated by improving the quality of the sample. The overall phonon lifetime can be obtained by summing up the inverse phonon lifetimes for these scattering processes, and it can be given as[64, 65]

$$\frac{1}{\tau} = \frac{1}{\tau_U} + \frac{1}{\tau_M}. \tag{8}$$

where $\tau_M$ is related to the mass-difference, and the phonon–phonon Umklapp scattering ($\tau_U$) involves three-phonon ($\tau_3$) and four-phonon ($\tau_4$) scatterings. Thus, the phonon lifetime can be rewritten as

$$\frac{1}{\tau} = \frac{1}{\tau_3} + \frac{1}{\tau_4} + \frac{1}{\tau_M}. \tag{9}$$

Based on perturbation theory, the balance of the phonon population split between diffusive drift and scattering can be established by the steady-state phonon Boltzmann transport equation as[66]

$$v_\lambda \cdot \nabla n_\lambda = \left.\frac{\partial n_\lambda}{\partial t}\right|_s, \tag{10}$$

where $v_\lambda$ is the group velocity attributed to mode $\lambda$, and $n_\lambda$ is the phonon occupation number. Three-phonon, four-phonon, as well as mass-difference scattering processes can be included in the scattering term on the right side of Eq. (10). The detailed derivation has been given in Ref. [67]. Therefore, combined with Eqs. (1) and (10), the inverse phonon lifetime ($\tau_{3/4/M}$) can be evaluated based on the following expansion[42, 67, 68]:

$$\frac{1}{\tau_3} = \sum_{\lambda_1 \lambda_2} \left[ \frac{1}{2} \frac{n^0_{\lambda_1} n^0_{\lambda_2}}{n^0_\lambda} \Gamma^-_{\lambda \lambda_1 \lambda_2} + \frac{(1+n^0_{\lambda_1}) n^0_{\lambda_2}}{n^0_\lambda} \Gamma^+_{\lambda \lambda_1 \lambda_2} \right],$$

$$\frac{1}{\tau_4} = \sum_{\lambda_1 \lambda_2 \lambda_3} \left[ \frac{1}{6} \frac{n^0_{\lambda_1} n^0_{\lambda_2} n^0_{\lambda_3}}{n^0_\lambda} \Gamma^{--}_{\lambda \lambda_1 \lambda_2 \lambda_3} + \frac{1}{2} \frac{(1+n^0_{\lambda_1}) n^0_{\lambda_2} n^0_{\lambda_3}}{n^0_\lambda} \Gamma^{+-}_{\lambda \lambda_1 \lambda_2 \lambda_3} + \frac{1}{2} \frac{(1+n^0_{\lambda_1})(1+n^0_{\lambda_2}) n^0_{\lambda_3}}{n^0_\lambda} \Gamma^{++}_{\lambda \lambda_1 \lambda_2 \lambda_3} \right], \tag{11}$$

$$\frac{1}{\tau_M} = \sum_{\lambda_1} \Gamma_{\lambda \lambda_1},$$

where $(n^0_\lambda)^{-1}$ obeys a phonon frequency $\omega$ of the Bose-Einstein distribution, that is

$$1 + \frac{1}{n^0_\lambda} = e^{\frac{\hbar \omega}{k_B T}}. \tag{12}$$

Note that the relationship between $(n^0_\lambda)^{-1}$ and $(n^0_{\lambda_{1/2/3}})^{-1}$ should obey the energy conservation law. For example, the relationship between three- and four-phonon absorption processes can be obtained by

$$\begin{cases} 1 + \dfrac{1}{n^0_\lambda} = \left(1 + \dfrac{1}{n^0_{\lambda_1}}\right)\left(1 + \dfrac{1}{n^0_{\lambda_2}}\right), & \text{if } \varpi_\lambda = \varpi_{\lambda_1} + \varpi_{\lambda_2} \\ 1 + \dfrac{1}{n^0_\lambda} = \left(1 + \dfrac{1}{n^0_{\lambda_1}}\right)\left(1 + \dfrac{1}{n^0_{\lambda_2}}\right)\left(1 + \dfrac{1}{n^0_{\lambda_3}}\right), & \text{if } \varpi_\lambda = \varpi_{\lambda_1} + \varpi_{\lambda_2} + \varpi_{\lambda_3} \end{cases} \tag{13}$$

Moreover, $\Gamma_\lambda^\pm$ refers to the various absorption and emission processes (refer to Ref. [42] for further details). The expressions for $\Gamma_\lambda^\pm$ and $\Gamma_\lambda$ can be given by Fermi's golden rule[41, 42, 67, 68] as

$$\Gamma_{\lambda\lambda_1\lambda_2}^\pm = \frac{\pi\hbar}{4N}\left|V_{\pm\lambda\lambda_1\lambda_2}^{(3)}\right|^2 \Delta_\pm \frac{\delta(\omega_\lambda \pm \omega_{\lambda_1} - \omega_{\lambda_2})}{\omega_\lambda \omega_{\lambda_1} \omega_{\lambda_2}},$$
$$\Gamma_{\lambda\lambda_1\lambda_2\lambda_3}^{\pm\pm} = \frac{\pi\hbar}{4N}\left|V_{\pm\pm\lambda\lambda_1\lambda_2\lambda_3}^{(4)}\right|^2 \Delta_{\pm\pm} \frac{\delta(\omega_\lambda \pm \omega_{\lambda_1} \pm \omega_{\lambda_2} - \omega_{\lambda_3})}{\omega_\lambda \omega_{\lambda_1} \omega_{\lambda_2} \omega_{\lambda_3}},$$
(14)

where $V_{\pm\lambda}^{(3)}\left[V_{\pm\lambda}^{(4)}\right]$ can be derived by Eq. (4).

Recent advances in phonon transport modeling have made it possible to calculate the lifetime of each phonon mode quantitatively. Based on perturbation theory, Marzari *et al.* reported a strong temperature dependence of the phonon lifetime of graphene in the typical operating range of 100–500 K[69]. Bonini *et al.* further demonstrated that for free-standing graphene, the transverse and longitudinal acoustic phonons in the long-wavelength limit have finite lifetimes (about 5 ps). This was entirely due to the quadratic dependence of the dispersions with momentum for the out-of-plane phonon flexural modes in free-standing low-dimensional systems[70]. Furthermore, Qin *et al.* reported that optical branches in monolayer black phosphorene had a short phonon lifetime, resulting in a low lattice thermal conductivity[71]. A similar result is also obtained in monolayer TMDs[29, 72-74]. Interestingly, Gu *et al.* found that the phonon lifetime of monolayer WS$_2$ is much longer than that in MoS$_2$[72]. The main reason is that the huge atomic weight difference between W and S results in a larger phonon band gap than MoS$_2$, which in turn prevented the scattering between phonon and optical branches, resulting in a longer phonon lifetime[72]. Gülseren *et al.* revealed that transition metal atoms (Mo or W) drove phonon lifetime, whereas chalcogen atoms (S or Se) determined the group velocity spectrum[75].

Previous studies analyzing the phonon lifetime were mainly based on three-phonon scattering. However, the phonon anharmonicity and lattice thermal conductivity of many bulk materials were overestimated[76-78]. Recently, Ruan *et al.* proposed that the four-phonon scattering process plays a more important role than the three-phonon scattering process, since it affects the thermal transport of optical phonons and significantly reduces the thermal conductivity of the material[79]. Moreover, Li *et al.* evaluated the phonon lifetimes of monolayer and hydrogenated bilayer BAs associated with three- and four-phonon scatterings at 300 K[80], as shown in Fig. 4. The relative intensities between three- and four-phonon scattering processes change with temperature. The main reason is that the three-phonon lifetime is inversely proportional to temperature, whereas the four-phonon lifetime exhibits a $T^{-2}$ dependence. Although the three-phonon scattering process underestimates the phonon lifetime, the trend of the three-phonon lifetime at low frequencies is almost similar to that of the four-phonon lifetime. These results illustrate that three-phonon scattering processes dominate around room temperature. Apart from that, there are many other factors affecting phonon scattering, such as chiral phonons. Chiral phonons have pseudo-angular momenta[81] and chirality which could modify the symmetry constraints for scattering[82, 83]. As a result, these features may influence the phonon lifetime[84], which in turn affects the thermal resistance.

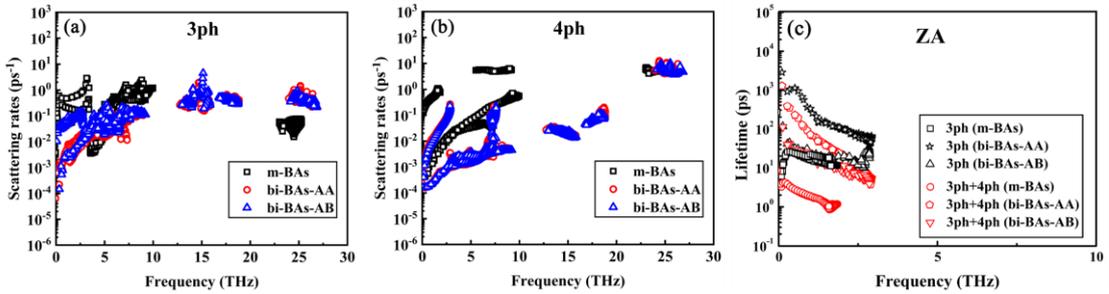

**Fig. 4** (a) Three-phonon and (b) four-phonon scattering rates as a function of frequency in three structures. (c) Phonon lifetime of ZA phonon branches as a function of frequency for those three structures[80]. Reproduced with permission. Copyright (2022), AIP Publishing.

Experimentally, measuring phonon lifetimes is quite difficult due to the stringent requirements placed on material samples and for obtaining a satisfying energy resolution. Generally, the energy–time uncertainty relation can be used to calculate phonon lifetimes from the phonon linewidth in Raman studies. To date, the phonon lifetimes of many 2D vdW materials, such as graphene[85-87], TMDs[88-90], and black phosphorene[45, 91], have been measured by Raman spectroscopy. While Raman scattering techniques can achieve resolutions in the micro-electron volt range, corresponding to resolved lifetimes up to nanoseconds, these measurements only detect the optical mode of the zone center[92]. Moreover, inelastic X-ray scattering techniques can detect phonon modes throughout the Brillouin zone, but they are limited to an energy resolution of 1-2 meV[93], which is insufficient to address the picosecond lifetime problem.

**C. Phonon frequency shift**

The shift of the phonon frequency is induced by third-order and higher-order IFCs according to Eq. (4). An example of an event leading to a phonon frequency shift is an exposure of the materials to a strain field since strain could alter the elastic constants, resulting in a change in thermal conductivity[94, 95]. Moreover, the temperature-dependent effect is also an issue affecting phonon frequency shift. At elevated temperatures, the higher-order terms in Eq. (2) cannot be neglected, and the anharmonic properties will become more and more prominent with increasing temperature. In other words, the atoms will no longer remain at their equilibrium positions at finite temperatures, but instead vibrate around them[96]. As a result, the force constants do not remain at their harmonic values and typically shift down. Therefore, the phonon frequencies are affected by an anharmonic effect and are lower than their harmonic values at 0K with increasing temperature. In this case, taking silicon as an example, Feng *et al.* rigorously calculated the phonon frequency shift caused by three and four-phonon scattering[97]. The results showed that the calculated phonon frequency shift was in good agreement with the inelastic neutron scattering data in the Brillouin region when fourth-order IFCs were considered as the fifth-nearest neighbor. Gu *et al.* demonstrated that four-phonon scattering in monolayer graphene was heavily affected by the temperature[98]. The effective anharmonic terms could be altered at high temperature.

## III. Thermal conductivity

2D vdW materials have broad application prospects in microelectronic devices such as field-effect transistors[32, 99, 100], diodes[101-103], and basic logic circuits[104-106]. One of the key issues in the design of these devices is how to dissipate heat efficiently, because in many cases, smaller electronic components generate a higher density of excess heat. Efficient heat dissipation depends on the material having a high thermal conductivity. The thermal conductivity of a material is composed of the electron thermal conductivity ($\kappa_e$) and lattice thermal conductivity ($\kappa_l$), as both electrons and phonons are heat transport carriers in solids. Electrons are the primary heat carriers in metallic materials, whereas phonons are the primary heat carriers in semiconductors. However, even in 2D vdW metallic materials, phonons play a major role in thermal transport, which is related to the quantum confinement effect[64, 107]. Here, only the lattice thermal conductivity of 2D vdW materials is discussed; at a given temperature $T$, the lattice thermal conductivity can be given as[108, 109]

$$\kappa_l(T) = \sum \tau C_v v^2. \tag{15}$$

where $C_v$ and $v$ are the specific heat and phonon group velocity, respectively. From Eq. (15), it is not difficult to see that phonon anharmonicity (related to the lifetime $\tau$) strongly affects the lattice thermal conductivity.

In 2D vdW materials, interfaces play a significant role in leading to different thermal transport mechanisms from their bulk materials, *via* the formation of new phonon bands or by inducing interfacial thermal resistance. Available theoretical and experimental data involving the thermal conductivity of typical 2D vdW materials at 300 K supports the thickness-induced effect (see Table I). Moreover, the thermal conductivity of 2D vdW materials exhibits a strong temperature dependence and declines gradually in the presence of a temperature gradient (see Fig. 5), which is mainly induced by the suppression of the phonon lifetime. However, this expectation does not apply in the case of monolayer graphene, a gapless semimetal, which exhibits a high thermal conductivity of up to 5300 Wm$^{-1}$K$^{-1}$[110]. Furthermore, Lindsay *et al.* investigated the thermal conductivity of graphene by considering three-phonon scattering[111]. They identified that due to the reflection symmetry in graphene, the out-of-plane ZA phonon branch had a significantly long relaxation time and contributed to over 70% of the total thermal conductivity while the scattering of ZA is symmetrically prohibited. However, Feng *et al.* demonstrated that the relative contribution of the ZA phonon branch to the thermal conductivity was reduced from 70% to 30% when four-phonon scattering was introduced[67], which fits better with the results based on MD simulations[112-114]. The main reason was that four-ZA processes ( ZA ⇌ ZA+ZA+ZA and ZA+ZA ⇌ ZA+ZA ) are still activated in the scattering process even in the presence of the reflection symmetry in graphene. This conclusion also resolves a long-debated problem pertaining to the extent of the contribution by ZA modes to the thermal conductivity of graphene. In addition to typical 2D materials, other graphene-like metal oxide monolayers have also been studied, such as BeO[115] with a thermal conductivity of about 266 Wm$^{−1}$K$^{−1}$ at 300 K and IIA–VI monolayer monoxides[116].

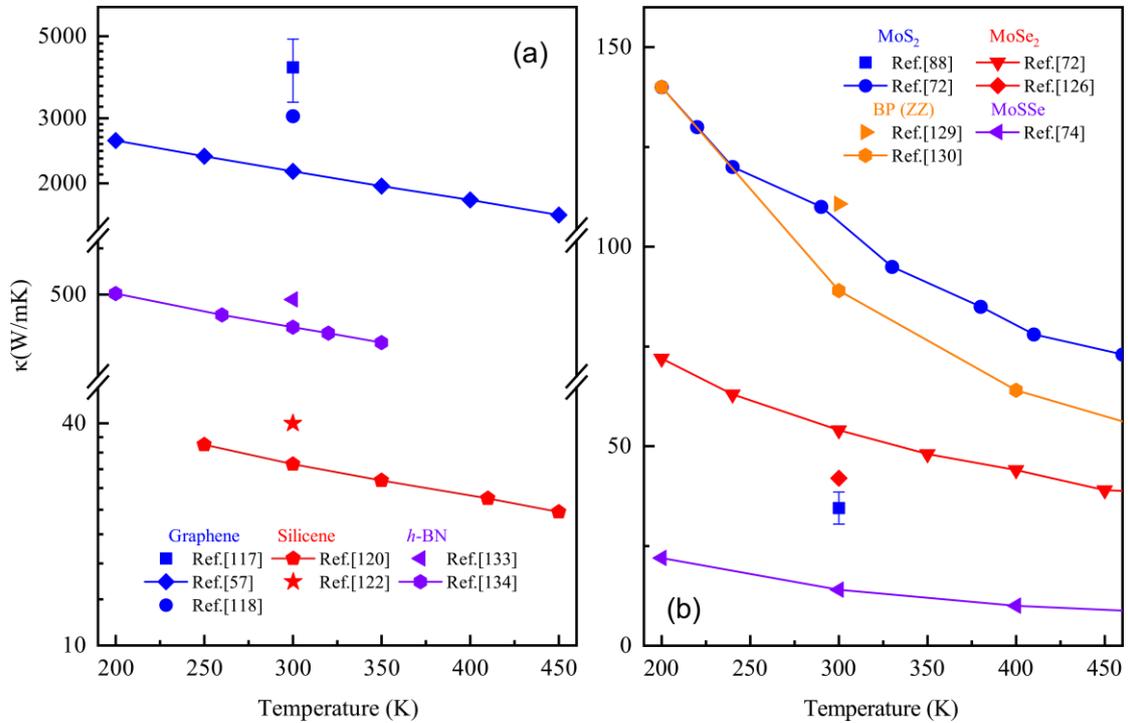

**Fig. 5** Temperature-dependent thermal conductivity of typical 2D vdW materials.

Single points refer to the thermal conductivity of the material measured experimentally at 300 K[57, 72, 74, 88, 117, 118, 120, 122, 126, 129, 130, 133, 134]. Reproduced with permission.

**Table I** Thermal conductivity of representative 2D vdW materials at room temperature (300 K) [57, 72, 74, 88, 117-134].

| Materials | | Dimension | Type | Method | $\kappa$ (300 K) [Wm$^{-1}$K$^{-1}$] | Reference |
|---|---|---|---|---|---|---|
| Graphene | | Monolayer | 2H | MRS | 3080 - 5150 | Ref. [117] |
| | | | | First-principles calculations | 2200 | Ref. [57] |
| | | | | MD simulation | 3038 | Ref. [118] |
| | | Bilayer | 2H | SCC-DFTB | 2748 | Ref. [119] |
| Silicene | | Monolayer | 2H | First-principles calculations | 32 | Ref. [120] |
| | | | | | 26 | Ref. [121] |
| | | | | MD simulation | 40 | Ref. [122] |
| TMDs | MoS$_2$ | Monolayer | 1T | MD simulation | 32 ± 3 | Ref. [123] |
| | | Monolayer | 2H | MRS | 34.5 ± 4 | Ref. [88] |
| | | | | First-principles calculations | 103 | Ref. [72] |
| | WS$_2$ | Monolayer | 2H | MRS | 32 | Ref. [124] |
| | | | | First-principles calculations | 142 | Ref. [72] |
| | MoSe$_2$ | Monolayer | 2H | OTR | 59 ± 18 | Ref. [125] |
| | | | | First-principles calculations | 54 | Ref. [72] |
| | | | | MD simulation | 42.3(AC) 42.0(ZZ) | Ref. [126] |
| | WSe$_2$ | Monolayer | 2H | OTR | 37 ± 12 | Ref. [127] |
| | | | | First-principles calculations | 53 | Ref. [72] |
| | MoSSe | Monolayer | 2H | First-principles calculations | 13.5 | Ref. [74] |
| | WSSe | Monolayer | 2H | First-principles calculations | 11 | Ref. [128] |
| Black phosphorene | | Monolayer | 2H | First-principles calculations | 24.3(AC) 83.5(ZZ) | Ref. [129] |
| | | | | MD simulation | 63.6(AC) 110.7(ZZ) | Ref. [130] |
| InSe | | Monolayer | $\gamma$ phase | First-principles calculations | 41.7 | Ref. [131] |
| | | | | First-principles calculations | 63.7 | Ref. [132] |

| | | | Microbridge | 484 | Ref. [133] |
|---|---|---|---|---|---|
| h-BN | Monolayer | 2H | | | |
| | | | First-principles calculations | 520 | Ref. [134] |

AC: Armchair edge

ZZ: Zigzag edge

SCC-DFTB: Self-consistent charge density functional tight binding

MRS: Micro-Raman spectroscopy

OTR: Optothermal Raman technique

The size effect of 2D vdW materials is not negligible in microelectronics, as these materials always require appropriate substrate support and integration. Earlier studies reported that the thermal conductivity is length-dependent for graphene patches with lengths ranging from 1 to 10 μm and it has a slow divergence[111]. Wei *et al.* demonstrated that the thermal conductivity of single-layer graphene also increases almost linearly with the sample length in the nanometre regime (ranging from 7 nm to 25 nm)[135]. Chen *et al.* further predicted that the thermal conductivity increases continuously without convergence for single-layer graphene lengths reaching up to 200 nm[136]. To quantitatively analyze the size effect on the thermal conductivity of materials, an analytical formulation to describe the size-confinement effect of phonon transport was proposed by Li *et al.*[137]. Although the size dependence of thermal conductivity was not observed by the optothermal Raman technique[110, 138], Xu *et al.* employed direct thermal-bridge measurements to discern that the room-temperature thermal conductivity of graphene kept increasing and was divergent with sample length following a logarithmical law[139]. Moreover, a size-dependent thermal conductivity was also found in other 2D vdW materials[72, 121, 129, 140-142].

Finally, we note that there are numerous studies and reviews on different aspects of the thermal properties of nanostructured materials. Comprehensive summaries are available in Ref. [143-148] and recent progress on nanoscale heat transport experiments is discussed in Ref. [23, 149, 150]. There are also comments on anomalous and exotic heat transport behaviors in low-dimensional materials[65, 66, 151-153].

**IV. Summary**

In this review, the progress of our understanding on phonon anharmonicity and the thermal conductivity of 2D vdW materials is discussed. Some important parameters for elucidating phonon scattering processes and thermal properties are reviewed and analyzed including the Grüneisen parameter and phonon lifetime. The derivation of phonon anharmonicity and the typical negative Grüneisen parameter in 2D vdW materials are illustrated systematically. The origins and evolution of three-phonon and four-phonon scattering processes that lead to finite phonon lifetimes are thoroughly examined. Furthermore, the relationship between the phonon frequency shift and temperature is expounded. We have also briefly compiled the size-dependent thermal conductivity of representative 2D vdW materials from both experimental and theoretical perspectives.

**Acknowledgments**

This work was supported by the 100 Talents Program of Sun Yat-Sen University (Grant 76220-18841201), the National Natural Science Foundation of China (Grant 22022309), and the Natural Science Foundation of Guangdong Province, China (2021A1515010024), the University of Macau (SRG2019-00179-IAPME, MYRG2020-00075-IAPME), and the Science and Technology Development Fund from Macau SAR (FDCT-0163/2019/A3).

# Reference

1. P. R. Wallace, *Phys. Rev.* **71**, 622-634 (1947).
2. K. S. Novoselov, A. K. Geim, S. V. Morozov *et al.*, *Sci.* **306**, 666-669 (2004).
3. Y. Zhang, Y.-W. Tan, H. L. Stormer *et al.*, *Nature* **438**, 201-204 (2005).
4. M. D. Stoller, S. Park, Y. Zhu *et al.*, *Nano Lett.* **8**, 3498-3502 (2008).
5. R. R. Nair, P. Blake, A. N. Grigorenko *et al.*, *Sci.* **320**, 1308-1308 (2008).
6. J. Qiao, X. Kong, Z.-X. Hu *et al.*, *Nat. Commun.* **5**, 4475 (2014).
7. K. F. Mak, C. Lee, J. Hone *et al.*, *Phys. Rev. Lett.* **105**, 136805 (2010).
8. A. Kuc, N. Zibouche, T. Heine, *Phys. Rev. B* **83**, 245213 (2011).
9. Q. H. Wang, K. Kalantar-Zadeh, A. Kis *et al.*, *Nat. Nanotechnol.* **7**, 699-712 (2012).
10. D. Akinwande, N. Petrone, J. Hone, *Nat. Commun.* **5**, 5678 (2014).
11. P. Avouris, *Nano Lett.* **10**, 4285-4294 (2010).
12. H. Jang, Y. J. Park, X. Chen *et al.*, *Adv. Mater.* **28**, 4184-4202 (2016).
13. S. Barraza-Lopez, B. M. Fregoso, J. W. Villanova *et al.*, *Rev. Mod. Phys.* **93**, 011001 (2021).
14. M. Donarelli, L. Ottaviano, *Sensors* **18**, 3638 (2018).
15. M. Xu, T. Liang, M. Shi *et al.*, *Chem. Rev.* **113**, 3766-3798 (2013).
16. A. K. Geim, I. V. Grigorieva, *Nature* **499**, 419-425 (2013).
17. J. Wu, J. Peng, H. Sun *et al.*, *Adv. Mater.* **n/a**, 2200425 (2022).
18. S. Zhang, S. Guo, Z. Chen *et al.*, *Chem. Soc. Rev.* **47**, 982-1021 (2018).
19. A. Carvalho, M. Wang, X. Zhu *et al.*, *Nat. Rev. Mater.* **1**, 16061 (2016).
20. S. Venkateshalu, G. Subashini, P. Bhardwaj *et al.*, *J. Energy Storage* **48**, 104027 (2022).
21. G. Mahan, B. Sales, J. Sharp, *Phys. Today* **50**, 42-47 (1997).
22. J. DiSalvo Francis, *Sci.* **285**, 703-706 (1999).
23. D. G. Cahill, W. K. Ford, K. E. Goodson *et al.*, *J Appl. Phys.* **93**, 793-818 (2002).
24. D. L. Nika, S. Ghosh, E. P. Pokatilov *et al.*, *Appl. Phys. Lett.* **94**, 203103 (2009).
25. Z. Y. Ong, E. Pop, *Phys. Rev. B* **84**, 075471 (2011).
26. J. H. Lan, J. S. Wang, C. K. Gan *et al.*, *Phys. Rev. B* **79**, 115401 (2009).
27. H. B. Zhou, Y. Q. Cai, G. Zhang *et al.*, *Phys. Rev. B* **94**, 045423 (2016).
28. Y. D. Kuang, L. Lindsay, S. Q. Shi *et al.*, *Nanoscale* **8**, 3760-3767 (2016).
29. Y. Q. Cai, J. H. Lan, G. Zhang *et al.*, *Phys. Rev. B* **89**, 035438 (2014).
30. X. J. Liu, G. Zhang, Q. X. Pei *et al.*, *Appl. Phys. Lett.* **103**, 133113 (2013).
31. J. W. Jiang, H. S. Park, T. Rabczuk, *J Appl. Phys.* **114**, 064307 (2013).
32. Z. Q. Fan, Z. H. Zhang, S. Y. Yang, *Nanoscale* **12**, 21750-21756 (2020).
33. X. Gu, C. Y. Zhao, *Comput. Mater. Sci.* **165**, 74-81 (2019).
34. X. Gu, R. Yang, *Phys. Rev. B* **94**, 075308 (2016).
35. J. Menéndez, M. Cardona, *Phys. Rev. B* **29**, 2051-2059 (1984).
36. R. A. Cowley, *Rep. Prog. Phys.* **31**, 123-166 (1968).
37. S. Narasimhan, D. Vanderbilt, *Phys. Rev. B* **43**, 4541-4544 (1991).
38. A. Shukla, M. Calandra, M. d'Astuto *et al.*, *Phys. Rev. Lett.* **90**, 095506 (2003).
39. A. Debernardi, S. Baroni, E. Molinari, *Phys. Rev. Lett.* **75**, 1819-1822 (1995).
40. A. A. Maradudin, A. E. Fein, *Phys. Rev.* **128**, 2589-2608 (1962).
41. A. Ward, D. A. Broido, D. A. Stewart *et al.*, *Phys. Rev. B* **80**, 125203 (2009).
42. T. Feng, X. Ruan, *Phys. Rev. B* **93**, 045202 (2016).

43. D. J. Ecsedy, P. G. Klemens, *Phys. Rev. B* **15**, 5957-5962 (1977).
44. L. Paulatto, F. Mauri, M. Lazzeri, *Phys. Rev. B* **87**, 214303 (2013).
45. D. Tristant, A. Cupo, X. Ling *et al.*, *ACS Nano* **13**, 10456-10468 (2019).
46. C. Carbogno, R. Ramprasad, M. Scheffler, *Phys. Rev. Lett.* **118**, 175901 (2017).
47. L. Lindsay, D. A. Broido, T. L. Reinecke, *Phys. Rev. B* **87**, 165201 (2013).
48. S. Li, Z. Tong, H. Bao, *J Appl. Phys.* **126**, 025111 (2019).
49. C. Cazorla, R. Rurali, *Phys. Rev. B* **105**, 104401 (2022).
50. Z. Tong, A. Pecchia, C. Yam *et al.*, *Adv. Funct. Mater.* **n/a**, 2111556 (2022).
51. E. Liang, Q. Sun, H. Yuan *et al.*, *Front. Phys.* **16**, 53302 (2021).
52. T. Peña, A. Azizimanesh, L. Qiu *et al.*, *J Appl. Phys.* **131**, 024304 (2022).
53. F. D. Stacey, J. H. Hodgkinson, *Phys. Earth Planet. Inter.* **286**, 42-68 (2019).
54. N. Mounet, N. Marzari, *Phys. Rev. B* **71**, 205214 (2005).
55. T. H. K. Barron, J. G. Collins, G. K. White, *Adv. Phys.* **29**, 609-730 (1980).
56. T. M. G. Mohiuddin, A. Lombardo, R. R. Nair *et al.*, *Phys. Rev. B* **79**, 205433 (2009).
57. B. D. Kong, S. Paul, M. B. Nardelli *et al.*, *Phys. Rev. B* **80**, 033406 (2009).
58. H. Sun, G. Liu, Q. Li *et al.*, *Phys. Lett. A* **380**, 2098-2104 (2016).
59. Y. Aierken, D. Çakır, C. Sevik *et al.*, *Phys. Rev. B* **92**, 081408 (2015).
60. L. Wang, C. Wang, Y. Chen, *J Phys: Condens. Mat.* **31**, 465003 (2019).
61. G. Liu, J. Zhou, *J Phys: Condens. Mat.* **31**, 065302 (2018).
62. A. M. Chippindale, S. J. Hibble, E. Marelli *et al.*, *Dalton Trans.* **44**, 12502-12506 (2015).
63. G. Qin, Z. Qin, W.-Z. Fang *et al.*, *Nanoscale* **8**, 11306-11319 (2016).
64. J. Zou, A. Balandin, *J Appl. Phys.* **89**, 2932-2938 (2001).
65. A. M. Marconnet, M. A. Panzer, K. E. Goodson, *Rev. Mod. Phys.* **85**, 1295-1326 (2013).
66. H. Bao, J. Chen, X. Gu *et al.*, *ES Energy environ.* **1**, 16-55 (2018).
67. T. Feng, X. Ruan, *Phys. Rev. B* **97**, 045202 (2018).
68. L. Lindsay, D. A. Broido, *J Phys: Condens. Mat.* **20**, 165209 (2008).
69. N. Bonini, M. Lazzeri, N. Marzari *et al.*, *Phys. Rev. Lett.* **99**, 176802 (2007).
70. N. Bonini, J. Garg, N. Marzari, *Nano Lett.* **12**, 2673-2678 (2012).
71. L. Yu, Y. Tian, X. Zheng *et al.*, *Int. J. Therm. Sci.* **174**, 107438 (2022).
72. X. Gu, R. Yang, *Appl. Phys. Lett.* **105**, 131903 (2014).
73. A. Kandemir, H. Yapicioglu, A. Kinaci *et al.*, *Nanotechnology* **27**, 055703 (2016).
74. S. D. Guo, *Phys. Chem. Chem. Phys.* **20**, 7236-7242 (2018).
75. A. Mobaraki, C. Sevik, H. Yapicioglu *et al.*, *Phys. Rev. B* **100**, 035402 (2019).
76. K. Esfarjani, G. Chen, H. T. Stokes, *Phys. Rev. B* **84**, 085204 (2011).
77. L. Lindsay, D. A. Broido, T. L. Reinecke, *Phys. Rev. Lett.* **111**, 025901 (2013).
78. L.-D. Zhao, S.-H. Lo, Y. Zhang *et al.*, *Nature* **508**, 373-377 (2014).
79. X. Yang, T. Feng, J. Li *et al.*, *Phys. Rev. B* **100**, 245203 (2019).
80. C. Yu, Y. Hu, J. He *et al.*, *Appl. Phys. Lett.* **120**, 132201 (2022).
81. L. Zhang, Q. Niu, *Phys. Rev. Lett.* **115**, 115502 (2015).
82. T. Pandey, C. A. Polanco, V. R. Cooper *et al.*, *Phys. Rev. B* **98**, 241405 (2018).
83. Q. Wang, S. Li, J. Zhu *et al.*, *Phys. Rev. B* **105**, 104301 (2022).
84. C. P. Romao, *Phys. Rev. B* **100**, 060302 (2019).
85. I. Calizo, A. A. Balandin, W. Bao *et al.*, *Nano Lett.* **7**, 2645-2649 (2007).
86. A. C. Ferrari, J. C. Meyer, V. Scardaci *et al.*, *Phys. Rev. Lett.* **97**, 187401 (2006).

87. S. Claramunt, A. Varea, D. López-Díaz *et al.*, *J Phys. Chem. C* **119**, 10123-10129 (2015).
88. R. Yan, J. R. Simpson, S. Bertolazzi *et al.*, *ACS Nano* **8**, 986-993 (2014).
89. P. Soubelet, A. E. Bruchhausen, A. Fainstein *et al.*, *Phys. Rev. B* **93**, 155407 (2016).
90. M. M. Petrić, M. Kremser, M. Barbone *et al.*, *Phys. Rev. B* **103**, 035414 (2021).
91. Z. Guo, H. Zhang, S. Lu *et al.*, *Adv. Funct. Mater.* **25**, 6996-7002 (2015).
92. A. M. A. Leguy, A. R. Goñi, J. M. Frost *et al.*, *Phys. Chem. Chem. Phys.* **18**, 27051-27066 (2016).
93. A. Q. R. Baron, SpringerNature, Switzerland, pp. 1643-1719, (2016), 10.1007/978-3-319-14394-1.
94. R. C. Picu, T. Borca-Tasciuc, M. C. Pavel, *J Appl. Phys.* **93**, 3535-3539 (2003).
95. W.-P. Hsieh, M. D. Losego, P. V. Braun *et al.*, *Phys. Rev. B* **83**, 174205 (2011).
96. T. Feng, B. Qiu, X. Ruan, *J Appl. Phys.* **117**, 195102 (2015).
97. T. Feng, X. Yang, X. Ruan, *J Appl. Phys.* **124**, 145101 (2018).
98. X. Gu, Z. Fan, H. Bao *et al.*, *Phys. Rev. B* **100**, 064306 (2019).
99. X. F. Yan, Q. Chen, L. L. Li *et al.*, *Nano Energy* **75**, 104953 (2020).
100. M. Long, A. Gao, P. Wang *et al.*, *Sci. Adv.* **3**, e1700589 (2017).
101. B. Ezhilmaran, A. Patra, S. Benny *et al.*, *J. Mater. Chem C* **9**, 6122-6150 (2021).
102. G. Collins Philip, A. Zettl, H. Bando *et al.*, *Sci.* **278**, 100-102 (1997).
103. J. Jadwiszczak, J. Sherman, D. Lynall *et al.*, *ACS Nano* **16**, 1639-1648 (2022).
104. B. Radisavljevic, M. B. Whitwick, A. Kis, *ACS Nano* **5**, 9934-9938 (2011).
105. A. Dathbun, Y. Kim, Y. Choi *et al.*, *ACS Appl. Mater. Inter.* **11**, 18571-18579 (2019).
106. D. Panigrahi, R. Hayakawa, Y. Wakayama, *J. Mater. Chem C* **10**, 5559-5566 (2022).
107. A. Balandin, K. L. Wang, *Phys. Rev. B* **58**, 1544-1549 (1998).
108. W. S. Capinski, H. J. Maris, T. Ruf *et al.*, *Phys. Rev. B* **59**, 8105-8113 (1999).
109. J. X. Cao, X. H. Yan, Y. Xiao *et al.*, *Phys. Rev. B* **67**, 045413 (2003).
110. A. A. Balandin, S. Ghosh, W. Bao *et al.*, *Nano Lett.* **8**, 902-907 (2008).
111. L. Lindsay, D. A. Broido, N. Mingo, *Phys. Rev. B* **82**, 115427 (2010).
112. D. L. Nika, A. A. Balandin, *J Phys: Condens. Mat.* **24**, 233203 (2012).
113. M. An, Q. Song, X. Yu *et al.*, *Nano Lett.* **17**, 5805-5810 (2017).
114. T. Feng, X. Ruan, Z. Ye *et al.*, *Phys. Rev. B* **91**, 224301 (2015).
115. C. Xia, W. Li, D. Ma *et al.*, *Nanotechnology* **31**, 375705 (2020).
116. C. Xia, Y. Zhao, D. Ma *et al.*, *J Phys: Condens. Mat.* **33**, 065701 (2020).
117. S. Ghosh, I. Calizo, D. Teweldebrhan *et al.*, *Appl. Phys. Lett.* **92**, 151911 (2008).
118. R. Su, X. Zhang, *Appl. Therm. Eng.* **144**, 488-494 (2018).
119. Y. Kuang, L. Lindsay, B. Huang, *Nano Lett.* **15**, 6121-6127 (2015).
120. Y. Guo, S. Zhou, Y. Bai *et al.*, *J. Supercond. Nov. Magn.* **29**, 717-720 (2016).
121. X. Gu, R. Yang, *J Appl. Phys.* **117**, 025102 (2015).
122. M. Hu, X. Zhang, D. Poulikakos, *Phys. Rev. B* **87**, 195417 (2013).
123. B. Mortazavi, T. Rabczuk, *RSC Adv.* **7**, 11135-11141 (2017).
124. N. Peimyoo, J. Shang, W. Yang *et al.*, *Nano Res.* **8**, 1210-1221 (2015).
125. X. Zhang, D. Sun, Y. Li *et al.*, *ACS Appl. Mater. Inter.* **7**, 25923-25929 (2015).
126. Y. Hong, J. Zhang, X. C. Zeng, *J Phys. Chem. C* **120**, 26067-26075 (2016).
127. E. Easy, Y. Gao, Y. Wang *et al.*, *ACS Appl. Mater. Inter.* **13**, 13063-13071 (2021).
128. A. Patel, D. Singh, Y. Sonvane *et al.*, *ACS Appl. Mater. Inter.* **12**, 46212-46219 (2020).
129. L. Zhu, G. Zhang, B. Li, *Phys. Rev. B* **90**, 214302 (2014).
130. Y. Hong, J. Zhang, X. Huang *et al.*, *Nanoscale* **7**, 18716-18724 (2015).

131. T. Pandey, D. S. Parker, L. Lindsay, *Nanotechnology* **28**, 455706 (2017).
132. Z. Zeng, S. Li, T. Tadano *et al.*, *J Phys: Condens. Mat.* **32**, 475702 (2020).
133. C. Wang, J. Guo, L. Dong *et al.*, *Sci. Rep.* **6**, 25334 (2016).
134. L. Lindsay, D. A. Broido, *Phys. Rev. B* **84**, 155421 (2011).
135. Z. Wei, Z. Ni, K. Bi *et al.*, *Carbon* **49**, 2653-2658 (2011).
136. J. Chen, G. Zhang, B. Li, *Nanoscale* **5**, 532-536 (2013).
137. L. H. Liang, B. Li, *Phys. Rev. B* **73**, 153303 (2006).
138. S. Chen, Q. Wu, C. Mishra *et al.*, *Nat. Mater.* **11**, 203-207 (2012).
139. X. Xu, L. F. C. Pereira, Y. Wang *et al.*, *Nat. Commun.* **5**, 3689 (2014).
140. Z. Tian, K. Esfarjani, J. Shiomi *et al.*, *Appl. Phys. Lett.* **99**, 053122 (2011).
141. T. Yue, Y. Sun, Y. Zhao *et al.*, *Phys. Rev. B* **105**, 054305 (2022).
142. X. Gu, B. Li, R. Yang, *J Appl. Phys.* **119**, 085106 (2016).
143. N. Li, J. Ren, L. Wang *et al.*, *Rev. Mod. Phys.* **84**, 1045-1066 (2012).
144. N. Yang, X. Xu, G. Zhang *et al.*, *AIP Adv.* **2**, 041410 (2012).
145. X. Xu, J. Chen, B. Li, *J Phys: Condens. Mat.* **28**, 483001 (2016).
146. J.-W. Jiang, B.-S. Wang, J.-S. Wang *et al.*, *J Phys: Condens. Mat.* **27**, 083001 (2015).
147. X. Gu, Y. Wei, X. Yin *et al.*, *Rev. Mod. Phys.* **90**, 041002 (2018).
148. X. Gu, R. Yang, *Annu. rev. heat transf.* **19** (2016).
149. M. M. Rojo, O. C. Calero, A. F. Lopeandia *et al.*, *Nanoscale* **5**, 11526-11544 (2013).
150. D. G. Cahill, P. V. Braun, G. Chen *et al.*, *Appl. Phys. Rev.* **1**, 011305 (2014).
151. A. A. Balandin, *Nat. Mater.* **10**, 569-581 (2011).
152. S. Liu, X. F. Xu, R. G. Xie *et al.*, *Eur. Phys. J. B* **85**, 337 (2012).
153. X. Gu, Z. Fan, H. Bao, *J Appl. Phys.* **130**, 210902 (2021).